\documentclass[a4paper,12pt]{article}

\usepackage{amsmath,amssymb,amsthm}
\usepackage[totalwidth=16.5cm,totalheight=23.5cm]{geometry}
\usepackage{graphicx}
\usepackage{comment}

\newcommand{\R}{{\mathbb R}}

\newcommand{\Z}{{\mathbb Z}}

\newcommand{\cL}{{\mathcal L}}

\newcommand{\su}{\mathfrak {su}}

\newcommand{\Mb}{\overline{M}}
\newcommand{\pa}{\partial}
\newcommand{\Si}{\Sigma}

\newcommand{\Oc}{\mathcal{O}}
\newcommand{\W}{\wedge}
\newcommand{\eps}{\epsilon}
\newcommand{\id}{{\mathbb I}}

\newcommand{\be}{\begin{eqnarray}}

\newcommand{\ee}{\end{eqnarray}}

\begin{document}
 \pagestyle{plain}
\title{Asymptotically hyperbolic connections}
\author{Joel Fine${}^{(1)}$, Yannick Herfray${}^{(2)}$, Kirill Krasnov${}^{(2)}$ and Carlos Scarinci${}^{(2)}$\\ {}\\ 
{\small \it ${}^{(1)}$ D\'epartement de math\'ematiques, Universit\'e libre de Bruxelles, B-1050 Bruxelles, Belgique} 
\\
{\small \it ${}^{(2)}$ School of Mathematical Sciences, University of Nottingham, NG7 2RD, UK} }
\maketitle
\begin{abstract}\noindent
General Relativity in 4 dimensions can be equivalently described as a dynamical theory of ${\rm SO}(3)\sim {\rm SU}(2)$-connections rather than metrics. We introduce the notion of asymptotically hyperbolic connections, and work out an analog of the Fefferman--Graham expansion in the language of connections. As in the metric setup, one can solve the arising ``evolution'' equations order by order in the expansion in powers of the radial coordinate. The solution in the connection setting is arguably simpler, and very straightforward algebraic manipulations allow one to see how the obstruction appears at third order in the expansion. Another interesting feature of the connection formulation is that the ``counter terms'' required in the computation of the renormalised volume all combine into the Chern--Simons functional of the restriction of the connection to the boundary. As the Chern-Simons invariant is only defined modulo large gauge transformations, the requirement that the path integral over asymptotically hyperbolic connections is well-defined requires the cosmological constant to be quantised. Finally, in the connection setting one can deform the 4D Einstein condition in an interesting way, and we show that asymptotically hyperbolic connection expansion is universal and valid for any of the deformed theories. 
\end{abstract}

\section{Introduction}

A ``conformally compact metric'' is a complete metric on the interior $M$ of a compact manifold with boundary $\Mb$, which can be conformally extended to the whole of $\Mb$ (see below for a more precise definition). The boundary $\partial \Mb$ inherits a natural conformal structure which one thinks of as the ``conformal infinity'' of the metric on $M$. 
An important subclass of conformally compact metrics are the so-called ``asymptotically hyperbolic metrics''. As the name suggests, these metrics behave as hyperbolic metrics near the conformal infinity. In particular, their sectional curvatures tend to a negative constant as one approaches the boundary. Special examples of such asymptotically hyperbolic metrics are the conformally compact metrics satisfying the Einstein condition, known in the mathematical literature as Poincar\'e--Einstein metrics.

A theory of Poincar\'e--Einstein manifolds was developed in an influential paper \cite{FG} by Fefferman and Graham (see also \cite{Fefferman:2007rka} for a more modern and detailed exposition). It was shown that a Poincar\'e--Einstein metric is determined by specifying a conformal metric on its boundary, together with an ``obstruction tensor'', and solving Einstein's equations order by order in powers of the ``radial'' coordinate. The Poincar\'e--Einstein metric is then given as an power series expansion, which is usually infinite. In particular, this construction serves the purpose of constructing conformal invariants (for the conformal metric at the boundary) from the more easily constructible local invariants of the associated Poincar\'e--Einstein metric on the interior. More recently, Poincar\'e--Einstein metrics have also played an important role in the physics literature, after papers \cite{Witten:1998qj}, \cite{Witten:1998zw} by Witten which highlight their role in the AdS/CFT correspondence. 

In four dimensions one can describe Riemannian geometry using ${\rm SO}(3)$ or ${\rm SU}(2)$-connections rather than metrics \cite{Fine:2013qta}. The parameterisation of metrics by connections is based on the fact that  a conformal structure is determined completely by the knowledge of which 2-forms are self-dual. One can then declare the three components of the curvature 2-forms of an ${\rm SU}(2)$-connection to be self-dual with respect to some metric. This defines a conformal metric, algebraically constructed from the curvature of the connection. There is also a particular representative in this conformal class of metrics that has a chance to satisfy the Einstein condition. For the convenience of the reader we give this connection description of Einstein metrics below.

The purpose of this paper is to develop the analog of the Fefferman--Graham expansion \cite{FG} in the language of connections. Unlike \cite{FG}, which works in any dimension, our setting utilises self-duality and thus only works in four dimensions. 

There are several motivations for this exercise. Firstly, one of the central differences between the parameterisations of geometry by metrics and connections is that in the latter case not all metrics can be obtained. By a simple count of the number of components, there are $3\times 4$ components in an ${\mathfrak su}(2)$-valued connection one-form. The metric that is constructed algebraically from the curvature 2-forms of this connection is a gauge-invariant object, and thus depends only on $12-3=9$ of the connection components. Thus, one may expect that arbitrary {\it conformal} metrics can be obtained from connections. However, unlike in Riemannian geometry,  one is no longer free to choose a representative in a given conformal class. To put it differently, if one views the space of Riemannian metrics as a fibre bundle over the space of conformal metrics, with the fibre being different metrics in the same conformal class, then the metrics that arise from connections give rise to a particular section of this fibre bundle (more precisely, a finite dimensional sub-bundle).

Directly related to this principal difference is the fact that, as we shall see below, in the connection setting we cannot talk separately about conformally compact and asymptotically hyperbolic conditions. Thus, connections with a prescribed behaviour near the boundary (analogues of conformally compact metrics) will automatically describe asymptotically hyperbolic metrics. This fact serves a motivation for the title of this paper. Whether this more restrictive setting is a drawback or an advantage very much depends on the context. Below we shall see that in situations when one would like to specialise to ``Einstein'' connections, i.e. analogues of metrics that satisfy Einstein equations, this will be an advantage because it will mean one less Einstein equation to solve.

The second motivation for our desire to reformulate the 4-dimensional Fefferman--Graham theory in terms of connections has to do with the notion of the renormalised volume. As was first described in \cite{Henningson:1998gx} in the general Fefferman--Graham setting, one can introduce a notion of the renormalised volume of a Poincar\'e--Einstein manifold as follows. Let $\rho$ be the defining function of our conformally compact manifold $(M,g)$, i.e. a function vanishing transversely at the boundary in such a way that $\rho^2 g$ can be extended to the boundary (see below for a more precise definition). One can then compute the volume of the manifold $M$ up to the surface $\rho=\epsilon$. As one sends $\epsilon\to 0$ the volume diverges, but the coefficients in front of the resulting inverse powers of $\epsilon$ are local invariants of the metric on the $\rho=\epsilon$ hypersurfaces. They can then be added as boundary terms to the volume functional/gravitational action so that, after these divergent contributions are removed, one obtains a finite renormalised volume. In even dimensions, and in particular in the case of 4D of interest for us here, the renormalised volume of a Poincar\'e--Einstein manifold is independent of the choice of $\rho$ and is thus an invariant of the metric in question.\footnote{In odd dimensions there is the so-called conformal anomaly, which makes the renormalised volume depend also on a representative of the conformal metric at infinity, see e.g.\ \cite{Krasnov:2006jb} for a description of the situation in 3 dimensions.}

In four dimensions there are two local boundary counter terms that need to be subtracted to get the renormalised volume. One of the nice features of the connection formulation to be developed here is that these two counter terms combine into the single term: the Chern--Simons invariant of the restriction of the ${\rm SU}(2)$-connection to the boundary. The renormalisation procedure is then stated and carried out in a more economical way. 

The appearance of the Chern-Simons invariant in renormalisation of the volume leads to the following interesting phenomenon. Let us assume that the conformal boundary is compact, as is common in the asymptotically hyperbolic setting. As is well-known, the Chern-Simons invariant of an ${\rm SU}(2)$ connection on a compact 3-manifold is not invariant under so-called large gauge transformations. For this reason, one usually writes the Chern-Simons action as $k/4\pi$ times the integral of the Chern-Simons 3-form, where $k$ is the so-called level. Under large gauge transformations the integral of the Chern-Simons form changes by multiples of $8\pi^2$. If one wants the Chern-Simons theory path integral to be well-defined, one wants the phase of the exponent in $\exp{i S}$ to be defined modulo $2\pi i$. This requires the level $k$ to be quantised. 

In our setting of asymptotically hyperbolic connections on 4-manifolds, the appearance of the Chern-Simons invariant  makes the renormalised volume only defined modulo multiples of a certain quantity, see below. More generally, even prior to imposing any field equations, the appearance of the Chern-Simons invariant of the connection as the boundary term in the action, and the fact that the Chern-Simons invariant is not invariant under large gauge transformations, makes the gravity action in the connection setup only defined modulo a certain quantity. If one takes the connection formulation seriously and requires the asymptotically hyperbolic connections path integral to be well-defined, this requires the cosmological constant, or rather the associated Hubble length, to be quantised in terms of Planck length squared, see below for a formula. We find this quantisation of $\Lambda$ an intriguing consequence of the connection formulation of gravity. 

The third motivation for our asymptotically hyperbolic connections story is that, once reformulated in terms of connections, the 4-dimensional Einstein condition can be deformed. Thus, one can introduce an infinite-parametric family of theories closely resembling Einstein's theory of gravity in their properties. These theories were first described in \cite{Krasnov:2006du}; more recent descriptions based on the connection formulation are \cite{Krasnov:2011up}, \cite{Krasnov:2011pp}. The reference that is of most relevance for us here is the recent paper \cite{Herfray:2015fpa} that also describes these modified gravity theories using a parameterisation with certain auxiliary fields. 

As we shall see below, all of these ``deformed'' theories admit asymptotically hyperbolic connections as solutions. In fact, we shall see that it is most natural to analyse asymptotically hyperbolic connections as well as the asymptotic expansion of the field equations in the general setting of an arbitrary theory from the infinite-parameter family of gravity theories \cite{Krasnov:2006du}. Field equations of any of these theories can be solved order by order in the radial coordinate in precisely the same way as is done for the case of Einstein connections. In particular, the freedom in specifying a certain trace-free transverse symmetric tensor that appears at certain order in the expansion is exactly the same for any member of our class of modified theories, as we shall see below. This gives one more illustration of the statement that all modified theories from the class to be described below are close to GR in their properties. 

Our final motivation for this study is the fact that the Fefferman--Graham expansion can be viewed as an example of an initial value problem for gravity, with the role of time being played by the radial coordinate. The gravitational initial value problem in terms of connections is rather different from that in terms of metrics. At the same time, the ability to solve the field equations order by order in expansion in powers of the radial coordinate renders the asymptotically hyperbolic initial value problem tractable, to a large extent. This study can be viewed as preparation for a more non-trivial analysis of the general initial value problem in terms of connections. 

The paper is organised as follows. In Section \ref{sec:FH} we review the relevant facts about the Fefferman--Graham expansion. Section \ref{sec:theory} reviews the connection formulation of GR. It also explains how modified gravitational theories are obtained by deforming the condition on the auxiliary matrix appearing in the Lagrangian. In Section \ref{sec:ah-conn} we compute the self-dual part of the Levi-Civita connection for an asymptotically hyperbolic metric, and use the result to motivate our definition of asymptotically hyperbolic connections. We also check that the metric associated to an asymptotically hyperbolic connection is itself asymptotically hyperbolic. Section \ref{sec:PE-conn} is the central one. Here we solve the connection evolution equations order by order in the radial coordinate. The main idea here is to first solve for the expansion of the auxiliary matrix. This solution is simple and universal, i.e. does not depend on which particularly theory from our family of theories one considers. In particular, it is applicable to GR. We also state the initial value problem in terms of connections in this section, and describe the gauge-fixing that is used in solving the evolution equations. We discuss the renormalised volume in terms of connection in Section \ref{sec:renorm}. The fact that the expansion of the auxiliary matrix $M$ starts to deviate from the identity matrix at a rather high order allows one to see very easily that the relevant counter terms are given by the Chern-Simons functional. We conclude with a discussion. Some extra details on the Hamiltonian formulation in the language of connections are given in the Appendix.

\section{Preliminaries: the Fefferman--Graham expansion}
\label{sec:FH}

The material in this section is standard, and is reviewed here for the sake of comparison with what happens in the connection setting. 

\subsection{Conformally compact metrics}

We start with a more precise definition of the conformally compact metrics. We call $(M,g)$ a \emph{conformally compact Riemannian manifold} if $M$ is the interior of a compact manifold $\overline M$ with boundary $\partial\overline M=X$ and there exists a smooth function $\rho:\overline{M}\to\R_{\geq 0}$ and a Riemannian metric $\gamma_{(0)}$ on $X$ such that
\begin{itemize}
	\item $\rho|_{X}=0$;
	\item $d\rho\neq0$ at all points of $X$;
	\item $\rho^2 g$ can be extended continuously to $X$;
	\item $\rho^2 g|_{TX}=\gamma_{(0)}$.
\end{itemize}
As previously stated, this means that even though the metric $g$ is only defined in $M$ and goes to infinity as one approaches the boundary of $\overline M$, there exists a representative $\rho^2 g$ in its conformal class that extends smoothly to the boundary at infinity. A good example to have in mind is the hyperbolic space $\mathbb {H}^4$ consisting of the upper-half of $\R^4$ with metric 
\be
g=\frac{dx^2+dy^2+dz^2+dt^2}{t^2},
\ee 
and $\rho=t$. 

It is important to note that the defining function $\rho$ is non-unique: for any positive function $\lambda$ on $\overline{M}$, we may consider $\tilde\rho=\lambda\rho$ which has the same properties above with $\gamma_{(0)}$ replaced by $\tilde\gamma_{(0)}=\lambda^2\gamma_{(0)}$. Therefore a conformally compact metric on $M$ only defines a conformal structure on the boundary $X=\pa \Mb$. 

\subsection{Asymptotically hyperbolic metrics}

A hyperbolic metric is by definition a metric having a constant sectional curvature $-1/l^2$, where $l$ has units of length. By rescaling, one can always achieve $l=1$, but we will keep the factor of $l$ in all the formulas to make it easier to trace the dependence on the cosmological constant. The canonical example is again the hyperbolic space $\mathbb {H}^4$ itself.

A conformally compact metric $g$ is called \emph{asymptotically hyperbolic} if its sectional curvature is asymptotically $-1/l^2$. Obviously, $\mathbb {H}^4$ being conformally compact and hyperbolic is in particular asymptotically hyperbolic. The asymptotically hyperbolic condition is equivalent to $|d\rho|^2_{g}=\rho^2/l^2$ at $X$ as can be easily seen from the asymptotic behaviour of the Riemann tensor of a conformally compact metric:
\[ R_{\mu\nu\rho\sigma}=-\frac{|d\rho|_{g}^2}{\rho^2}(g_{\mu\rho}g_{\nu\sigma}-g_{\mu\sigma}g_{\nu\rho})+O(1/\rho). \]
We also see from this formula that conformally compact Einstein metrics, i.e.\ ones satisfying the Riemannian version of Einstein equations, are automatically asymptotically hyperbolic; they are called \emph{Poincar\'e--Einstein metrics}. From the above formula we also get the relation between the cosmological constant appearing in the Einstein condition $R_{\mu\nu}=\Lambda g_{\mu\nu}$ and $l$
\be
\Lambda = - \frac{3}{l^2}.
\ee

\subsection{Expansion}

Near the boundary $X$, we can write any conformally compact metric as
\be\label{cc-metric}
g=\frac{l^2}{ \rho^2}\Big(\alpha d\rho^2+d\rho\otimes\beta+\beta\otimes d\rho+\gamma\Big),
\ee
where $\alpha$, $\beta$ and $\gamma$ are 1-parameter families of functions, 1-forms and symmetric $(2,0)$-tensors on $X$, with the following power series expansion
\[ \alpha=1+\rho\alpha_{(1)}+\cdots,\qquad\beta=\beta_{(0)}+\rho\beta_{(1)}+\cdots,\qquad \gamma=\gamma_{(0)}+\rho\gamma_{(1)}+\cdots. \]
The asymptotically hyperbolic condition is $\alpha=1$. Furthermore, we can always achieve $\beta=0$ by an appropriate diffeomorphism: the gradient flow of $\rho$ with respect to $\rho^2 g$ identifies the level sets of $\rho$ and removes this mixed term. In the  asymptotically hyperbolic setting, we can thus write the metric in the form:
\be\label {FG}
g=\frac{l^2}{\rho^2}d\rho^2+\frac{l^2}{\rho^2}\gamma \qquad \gamma=\gamma_{(0)}+\rho\gamma_{(1)}+\cdots. 
\ee
There still remains some diffeomorphism freedom: infinitesimal diffeomorphisms generated by vector fields $\xi+\rho\lambda\partial_\rho\in\Gamma(T\overline{M})$ where the function $\lambda$ is $\rho$-independent and the vector field $\xi$ satisfies a certain condition. We thus have, essentially, a residual boundary diffeomorphism as well as freedom of local rescalings of the radial coordinate. The latter is equivalent to the freedom of conformal rescalings of the boundary metric.

Now, it is an exercise to write down Einstein's equations for (\ref{FG}) and solve them order by order in terms of an initially given value of $\gamma_{(0)}$. The first two terms are given by
\be\label{gamma-2}
\gamma_{(1)}=0,\qquad \gamma_{(2)}{}_{ab}=-R_{(0)}{}_{ab}+\frac{1}{4}R_{(0)}\gamma_{(0)}{}_{ab}
\ee
and the third remains undetermined but constrained to satisfy 
$$\gamma_{(0)}^{ab}\gamma_{(3)}{}_{ab}=0,\qquad \nabla_{(0)}{}^a\gamma_{(3)}{}_{ab}=0.$$
The remaining, higher-order terms, are then recursively determined algebraically from the previous terms. The unconstrained part of $\gamma_{(3)ab}$ is referred to as the {\it obstruction} by Fefferman-Graham.

We can interpret the expansion (\ref{FG}) as an instance of the initial value problem. Given the initial data on the boundary in the form of the boundary metric $\gamma_{(0)}$, as well as the constrained $\gamma_{(3)}$, which is best thought of as an initial momentum, we can propagate these data into the bulk, obtaining a Poincar\'e--Einstein metric, at least formally in a neighbourhood of the boundary. This is known as the \emph{Fefferman--Graham expansion}.

\subsection{Renormalised volume}

Using the tracelessness of $\gamma_{(3)}$, an easy calculation gives
\be\nonumber
\sqrt{{\rm det}(\gamma)} = \sqrt{{\rm det}(\gamma_{(0)})} \left( 1+\frac{1}{2} \rho^2 {\rm Tr}(\gamma_{(0)}^{-1} \gamma_{(2)})+ O(\rho^4) \right) = \sqrt{{\rm det}(\gamma_{(0)})} \left( 1-\frac{1}{8} \rho^2 R_{(0)}+ O(\rho^4)\right),
\ee
where we have used (\ref{gamma-2}) to get the second equality. The volume of the metric (\ref{FG}), up to a surface $\rho=\epsilon$ is then
\be
 V_\epsilon =  l^4 \int_\epsilon \frac{d\rho}{\rho^4} \int_{X_\epsilon} \sqrt{{\rm det}(\gamma_{(0)})} \left( 1-\frac{1}{8} \rho^2 R_{(0)}+ O(\rho^4)\right),
\ee
where ${X_\epsilon}$ is the surface $\rho=\epsilon$. Thus, the volume is given by
\be\nonumber
V_\epsilon = \frac{l^4 }{3\epsilon^3} \int_{X_\epsilon} \sqrt{{\rm det}(\gamma_{(0)})} - \frac{l^4 }{8\epsilon} \int_{X_\epsilon} \sqrt{{\rm det}(\gamma_{(0)})} R_{(0)} + {\rm finite\,\,\, terms},
\ee
and the renormalised volume is then defined as
\be\label{RV}
RV:=\lim_{\epsilon\to 0} \left(  V_\epsilon -\frac{l^4}{3\epsilon^3} \int_{X_\epsilon} \sqrt{{\rm det}(\gamma_{(0)})} + \frac{l^4}{8\epsilon} \int_{X_\epsilon} \sqrt{{\rm det}(\gamma_{(0)})} R_{(0)} \right).
\ee
We note that the renormalized volume is independent of the choice of representative $\gamma_{(0)}$ of the conformal class on $X$. In fact for another representative $\tilde \gamma_{(0)}=\lambda^{2}\gamma_{(0)}$, the corresponding defining function is $\tilde \rho=\lambda\rho$ for an appropriate extension of $\lambda$ from the boundary into the bulk, and $\tilde{\epsilon}=\lambda^2 \epsilon$. Then the formula (\ref{RV}) takes exactly the same form in terms of the tilded quantities, which shows that it does not matter which representative in the conformal class of the boundary is used to compute it. 

\section{Einstein connections}
\label{sec:theory}

In this section we review how Einstein metrics can be described in terms of ${\rm SO}(3)$ or ${\rm SU}(2)$-connections. We follow the recent description in \cite{Herfray:2015fpa}, where the reader is referred for more details. 

\subsection{The action functional}

Even though a ``pure connection'' description of non-zero scalar curvature Einstein metrics is possible, a description with a certain set of auxiliary fields is more useful in the present context. 

Let $A^i$ be an ${\rm SO}(3)$-connection on a 4-manifold $M$, and $F^i = dA^i + (1/2) \epsilon^{ijk} A^j\wedge A^k$ be its curvature 2-forms. Let $M^{ij}$ be a symmetric $3\times 3$ matrix of auxiliary fields. Let us assume for the moment that the manifold in question is compact. We will discuss the case of manifolds with boundary, relevant for this work, later. Consider the following action:
\be\label{action}
S[A,M] = \int_M {\rm Tr}( M F\wedge F).
\ee
We also impose a constraint on the matrix $M$, that it lie on a co-dimension one hypersurface with defining equation
\be\label{surface}
g(M)=0
\ee
in the space of symmetric $3\times 3$ matrices, determined by a real valued function $g$. Different hypersurfaces give different field equations; the hypersurface relevant to Einstein connections/metrics is given by
\be\label{g-EH}
g_{\rm E}(M) = {\rm Tr}(M^{-1}) - \Lambda,
\ee
where $4\Lambda$ is the scalar curvature. The claim is that the critical points of (\ref{action}), with $M$ restricted to lie on (\ref{surface}) with $g$ given by (\ref{g-EH}) give rise to Einstein metrics.

\subsection{Euler--Lagrange equations}

Varying the action (\ref{action}) with respect to the connection we get
\be\label{eq-A}
d_A \left( M^{ij} F^j \right) =0.
\ee
Here we assumed that the manifold is compact, so that the integration by parts necessary to derive this field equation is justified. We will consider the issue of boundary terms below. 

To get the equation for $M$, it is useful to impose the constraint (\ref{surface}) with the help of a Lagrange multiplier. To do this, we introduce an auxiliary volume form ${\mathcal V}$. We can then write
\be\label{X}
F^i\wedge F^j = X^{ij} \,{\mathcal V}.
\ee
Note that a different choice of $\mathcal V$ changes the matrix $X^{ij}$ by multiplication by a function. The matrix $X$ is thus defined intrinsically only up to scale.
We can now write the action as
\be
S[A,M,\mu] = \int_M \left( {\rm Tr}(M X) - \mu \, g(M) \right) {\mathcal V}.
\ee
The Euler-Lagrange equations for $M$ are then
\be\label{eq-M}
X = \mu \frac{\partial g}{\partial M}.
\ee
(Here, we write $\partial g/\partial M$ for the matrix defined by
\[
\frac{\partial g}{\partial M^{ij}} A^{ji} = dg\Big|_{M}(A)
\]
for all symmetric $A^{ij}$.) Equation (\ref{eq-M}) can be solved, together with $g(M)=0$, for $M(X)$. This can then be substituted to (\ref{eq-A}) to obtain the ``pure connection'' formulation of field equations, in which we obtain second-order PDE's for the connection. Alternatively, one can insert the solution $M(X)$ back into the Lagrangian to obtain the ``pure connection'' action principle. One can check that the Lagrangian arising this way is just the Legendre transform of the function $g(M)$. 

However, in many circumstances a more useful approach is not to solve for $M(X)$, but to interpret (\ref{eq-A}) as a first-order PDE on the matrix $M$. Solving it one can then find $X$ from (\ref{eq-M}) and thus obtain a set of first-order PDE's on the connection components. It is this strategy that will be followed below for the asymptotically hyperbolic connections. 

Let us illustrate all this for the Einstein case (\ref{g-EH}). In this case we can rewrite (\ref{eq-M}) as $M X = - \mu M^{-1}$. Taking the trace and using the condition $g_E(M)=0$ we obtain $\mu$. Thus, all in all for the Einstein case
\be\label{eq-M-gr}
X = \frac{1}{\Lambda} {\rm Tr}(M X) M^{-2},
\ee
which gives $X$ in terms of $M$, up to scale. (We can do no better, since $X$ is only defined up to scale, which was fixed by our choice of $\mathcal V$.)

\subsection{The metric}

Given a connection $A$ and symmetric matrix-valued function $M^{ij}$, we now describe how to define a Riemannian metric on the 4-manifold. When the connection satisfies the Euler-Lagrange equations (\ref{eq-A}), (\ref{eq-M-gr}), the corresponding metric will be Einstein (see \cite{Herfray:2015fpa} for the proof).

First, one constructs the conformal metric. This is the unique conformal structure that makes the triple of curvature 2-forms $F^i$ self-dual. Explicitly, 
\be
\sqrt{{\rm det(g)}} \, g_{\mu\nu} \sim \tilde{\epsilon}^{\alpha\beta\gamma\delta} \epsilon^{ijk} F^i_{\mu\alpha} F^j_{\nu\beta} F^k_{\gamma\delta}.
\ee
(Note this requires a non-degeneracy condition on the $F^i$, that the matrix $X^{ij}$ above in equation (\ref{X}) be definite.) Second, one fixes the volume form ${\mathcal V}_g$ of the metric to be 
\be\label{volume}
- 2\Lambda \, {\mathcal V}_g = {\rm Tr}(M F\wedge F).
\ee
With this choice of volume form, the action (\ref{action}) has the geometric interpretation of the total volume, up to overall scale.  The numerical factor in this formula is adjusted to agree with some conventions introduced later. 

\subsection{Deformed Einstein connections}

An interesting class of deformations of the Einstein condition is obtained by simply changing the codimension-one hypersurface that the matrix $M$ is constrained to lie on, in the action (\ref{action}). This corresponds to changing the function $g$ in the equation $g(M)=0$. As is mentioned above, the function that gives rise to Einstein connections is (\ref{g-EH}). Repacing $M$ by $M^{-1}$ in this function, setting $g_{inst}={\rm Tr}(M) - \Lambda$, turns out to give self-dual General Relativity, with solutions being half-flat Einstein metrics, see \cite{Herfray:2015fpa} for a discussion of this.  Another very interesting modification that can be considered is to set 
\be
g_{mod}(M) = {\rm det}(M) - \Lambda.
\ee
This gives a theory in many ways simpler than GR. In fact, this is the theory that was studied by one of us in \cite{Joel}, albeit in its``pure connection'' version without the matrix $M$.  A further family of modifications has been considered recently in \cite{Herfray:2015fpa}, and relies on parametrising the matrix $M^{-1}$ in terms of its trace and the trace free parts. In this way, one can have controllable modifications of GR. 

Details of these modified theories will not be important for us here, because we will be solving the arising equations in general for an arbitrary function $g(M)$. All our results below apply equally to the modified theories as well as to GR. 

In the deformed case, one continues to define the metric by the condition that the curvature 2-forms are self-dual. The volume form is then fixed to be a constant multiple of the form ${\rm Tr}(M F\wedge F)$, so that the action (\ref{action}) continues to receive the interpretation of the total volume of the space. 

\section{Asymptotically hyperbolic connections}
\label{sec:ah-conn}

To motivate our definition of an asymptotically hyperbolic connection, we first compute the self-dual part of the Levi-Civita connection for an asymptotically hyperbolic metric (\ref{FG}). 

\subsection{${\rm SO}(3)$-connection of an asymptotically hyperbolic metric}

Given an asymptotically hyperbolic 4-manifold $(M,g)$, we now compute the Levi-Civita connection on the bundle of self-dual 2-forms. First we choose a local coframe for $g$ adapted to the chosen conformal compactification
$$\theta^i=\frac{l}{\rho}e^i,\qquad\theta^4=\frac{l}{\rho}d\rho,$$
where $e^i$ denotes a 1-parameter family of local coframes for the 1-parameter family of metrics $\gamma$ on $X$. The spin connection is then computed by decomposing the structure equation as
$$\begin{cases}d\theta^i+\omega^i{}_j\wedge \theta^j+\omega^i{}_4\wedge \theta^4=0
\cr
d\theta^4+\omega^4{}_i\wedge \theta^i=0, \end{cases}$$
or, equivalently 
$$\begin{cases}
de^i-\frac{1}{\rho}d\rho\wedge e^i+\omega^i{}_j\wedge e^j+\omega^i{}_4\wedge d\rho=0
\cr
\omega^4{}_i\wedge e^i=0.\end{cases}$$
These equations can be readily solved in terms of the (co-)triad $e^i$ and the associated $\gamma$-compatible spin connection ${}^\gamma\omega^i{}_j$
$$\begin{cases}
\omega^i{}_j={}^\gamma\omega^i{}_j-\frac{1}{2}\Big(\gamma(\dot e^i,e^j)-\gamma(\dot e^j,e^i)\Big)d\rho
\cr
\omega^{4i}=\frac{1}{\rho}e^i-\frac{1}{2}\Big(\gamma(\dot e^i,e^j)+\gamma(\dot e^j,e^i)\Big)e^j.
\end{cases}$$
Here dot denotes the derivative with respect to $\rho$, and $\gamma(\dot e^i,e^j)$ is the metric product of 1-forms $\dot e^i$ and $e^j$. 

We may now express the $SU(2)$-connection $A^i$ in terms of the 1-parameter families of (co-)frames and spin connections on $X$ as
\be\label{A-metric}
A^i=\frac{1}{\rho}e^i+{}^\gamma\omega^i-\frac{1}{2}\Big(\gamma(\dot e^i,e^j)+\gamma(\dot e^j,e^i)\Big)e^j+\frac{1}{2}\epsilon^{ijk}\gamma(\dot e^j,e^k)d\rho,
\ee
where ${}^\gamma\omega^i=-(1/2)\epsilon^{ijk}\, {}^\gamma\omega^{jk}$. This will motivate our definition of asymptotically hyperbolic $SU(2)$-connections below.

Let us note that for an asymptotically hyperbolic metric that is also Einstein, equation (\ref{gamma-2}) is satisfied. The first of these conditions then implies that the third term in (\ref{A-metric}) starts with a term $O(\rho)$, so that the connection ${}^\gamma\omega^i$ compatible with $e^i$ is the only contribution at order $O(1)$. 

\subsection{Asymptotically hyperbolic ${\rm SO}(3)$-connections}

Let $M$ be the interior of a compact manifold $\overline{M}$ with boundary $\partial\overline M=X$. Let $P\to\overline M$ be an ${\rm SO}(3)$ or ${\rm SU}(2)$-bundle over $\overline M$ and denote by $P|_M\to M$ its restriction to $M$ and by $P|_X\to X$ its restriction to $X$. We require $P|_X$ to be abstractly isomorphic to the frame bundle of $X$.

We shall call an ${\rm SU}(2)$-connection $A^i$ on $P|_M$ \emph{asymptotically hyperbolic} if there exists a smooth function $\rho:\overline{M}\to\R_+$ and a smooth connection $\omega$ on $P$ defined on the whole of $\overline{M}$, such that
\begin{itemize}
 \item $\rho$ is a boundary defining function, i.e., $\rho|_{X}=0$ and $d\rho|_{X}\neq0$;
 \item The endomorphism-valued 1-form $\rho(A^i-\omega^i)$ extends smoothly to $X$;
 \item The restriction $\rho(A^i-\omega^i)|_{TX}=e^i_{(0)}$ is a coframing for $X$. Equivalently, $\rho(A-\omega)|_{TX} \in T^*X \otimes \mathfrak(P|_X)$ gives an isomorphism $TX \to \mathfrak{su}(P|_X)$.
\end{itemize}
The last condition here should is the analogue of the condition that a conformally compact metric $g$ should have $\rho^2g|_X$ a metric on $X$.

It is immediate that the definition does not depend on the choices of $\rho$ or $\omega$. Given an asymptotically hyperbolic connection, we can define a metric $\gamma_{(0)}$ on $X$ by declaring $e^i_{(0)}$to be an orthonormal coframe. This fixes an isomorphism $P|_X \to TX$. It is then convenient to take $\omega^i$ to be an arbitrary extension of the Levi-Civita connection $\omega_{(0)}^i$ of $\gamma_{(0)}$ on $P|_X$ to the whole of $P \to \overline{M}$. 

We will make these definitions of $\gamma_{(0)}$ and $\omega$ throughout. Thus, in general, an asymptotically hyperbolic ${\rm SU}(2)$-connection may be written as:
\be\label{as-hyperb-conn}
A^i=\frac{1}{\rho}e^i_{(0)}+\omega^i_{(0)}+a^i+b^id\rho
\ee
where
\[ a^i=a^i_{(0)}+\rho a^i_{(1)}+\cdots, \qquad b^i=b^i_{(0)}+\rho b^i_{(1)}+\cdots \]
are 1-parameter families of 1-forms and functions on $X$, respectively. 

\subsection{Metric of an asymptotically hyperbolic connection}

In this subsection we compute the metric associated to an asymptotically hyperbolic connection. We find it to be asymptotically hyperbolic. As we have already emphasised in the Introduction, one of the main differences between the metric and connection setups is that in the later case the requirement (\ref{as-hyperb-conn}) of a first-order pole in the radial coordinate automatically gives rise to connections asymptotically satisfying the field equations. This should be contrasted to the metric case, where the requirement (\ref{FG}) of a second-order pole only gives the conformal compactness condition and an extra condition must be imposed to obtain asymptotically hyperbolic metrics.

The metric is computed in two steps. First, we need to find a metric that makes the curvature 2-forms self-dual. The curvature of an asymptotically hyperbolic connection is:
\be\label{curv-asympt}
F^i = -\frac{1}{\rho^2}\Si^i_{(0)}+\Oc (1) = -\frac{1}{\rho^2}\left( d\rho\W e^i_{(0)}-\frac{\eps^{ijk}}{2}e^j_{(0)}\W e^k_{(0)} \right )+\Oc (1). 
\ee
Note that there is no $1/\rho$ term here because $d_{\omega_{(0)}} e_{(0)}^i = 0$ (since $\omega_{(0)}$ is the Levi-Civita connection of $\gamma_{(0)}$). 
The conformal metric in which these 2-forms are self-dual is a multiple of 
\[ 
\frac{1}{\rho^2}\left(d\rho^2+ \sum_{i} e^i_{(0)}\otimes e^i_{(0)}\right)+\Oc (1).
\]
As is clear from (\ref{curv-asympt}), asymptotically the matrix (\ref{X}) of wedge products $X^{ij}\sim\delta^{ij}$. From (\ref{eq-M-gr}) this implies that the matrix $M$ is asymptotically pure trace $M^{ij}\sim \delta^{ij}$. This in particular implies that the field equation (\ref{eq-A}) is automatically satisfied, as it reduces to the Bianchi identity for the curvature. Thus, asymptotically hyperbolic connections are automatically asymptotic solutions of the field equations. 

To fix the metric, we compute, asymptotically
\be
{\rm Tr}(M F\wedge F) = \frac{3}{\Lambda}{\rm Tr}(F\wedge F) = - \frac{3}{\Lambda} \frac{6}{\rho^4} d\rho\wedge e_{(0)}^1\wedge e_{(0)}^2\wedge e_{(0)}^3.
\ee
From (\ref{volume}), remembering $\Lambda=-3/l^2$, we immediately get that the metric defined by the connection, asymptotically, is
\be
\frac{l^2}{\rho^2}\left(d\rho^2+ \sum_{i} e^i_{(0)}\otimes e^i_{(0)}\right)+\Oc (1),
\ee
which is an asymptotically hyperbolic metric of sectional curvature $-1/l^2$. 

\subsection{Boundary conditions}

Let us now discuss the well-posedness of the connection variational principle in the asymptotically hyperbolic setup with boundary. The variation of ${\rm Tr}(M F\wedge F)$ produces a boundary term 
\be\label{b-term}
\delta \int {\rm Tr}(M F\wedge F) = 2 \int_X {\rm Tr}( M F \wedge \delta A) + \ldots,
\ee
where the dots stand for terms proportional to field equations. As we just saw, asymptotically $M^{ij}\sim \delta^{ij}$. A convenient choice of boundary conditions for the variational principle is then as follows. Instead of fixing the connection on the boundary, which is inconvenient because the connection has a first order pole there, we can require that asymptotically and to a sufficiently high order in the radial coordinate expansion (see below) the matrix of auxiliary fields is a multiple of the identity matrix. Then the boundary term (\ref{b-term}) arising in the variation of the action is cancelled by the Chern--Simons invariant of the restriction of the connection to the boundary
\be\label{CS}
S^{CS}_X[A] = \int_X \left( A^i \wedge dA^i + \frac{1}{3} \epsilon^{ijk} A^i\wedge A^j\wedge A^k\right),
\ee
which has the property
\be
\delta S^{CS}_X[A] = 2 \int_X {\rm Tr}(F \wedge \delta A).
\ee
Thus, with the boundary condition being those on $M$, the required boundary term is just the Chern--Simons invariant of the boundary connection. Below we shall see that this is also a good choice for cancelling the divergent contributions coming from the bulk part of the action, once everything is evaluated on a solution to field equations. 

The Chern-Simons invariant of an ${\rm SO}(3)$ connection is not gauge-invariant, being defined only modulo multiples of $8\pi^2$, see below. We will further discuss this ambiguity and its consequences below.

\section{Poincar\'e--Einstein connections}
\label{sec:PE-conn}

In this section we solve the field equations (\ref{eq-A}) and (\ref{eq-M}) by expanding the fields in a power series in the radial coordinate. We start by discussing the 3+1 decomposition of the connection field equations. 

\subsection{Initial value formulation}
\label{subsec: Init val form}

We fix a boundary defining function $\rho\in C^\infty(M,\R_{\geq 0})$ and decompose the connection $A^i$ as
$$A^i=\alpha^i+\beta^id\rho$$
where $\alpha^i$ and $\beta^i$ are 1-parameter families of 1-forms and functions on $X$. The curvature of $A$ is then
$$F_A^i=f_\alpha^i-(\dot\alpha^i-d^X_\alpha\beta^i)\wedge d\rho,$$
where $d^X_\alpha$ is the covariant exterior derivative with respect to connection $\alpha^i$ along the hypersurface $\rho=\text{const}$, and $f^i_\alpha$ is the curvature of $\alpha^i$. We may define the matrix $X^{ij}$ of curvature wedge products (up to scale) as
\be\label{X-alpha}
X^{ij} \sim f_\alpha^{(i}\wedge (\dot\alpha^{j)}-d^X_\alpha\beta^{j)})\wedge d\rho.
\ee

Defining the momentum $\pi^i$ canonically conjugate to $\alpha^i$ by
\be\label{pi}
\pi^i = M^{ij} f^j_\alpha,
\ee
the 3+1 decomposition of the field equation (\ref{eq-A}) reads
\be\label{feqs}
\begin{cases}
\dot\pi^i+[\beta,\pi]^i=d^X_\alpha\Big(M^{ij}(\dot\alpha^j-d^X_\alpha\beta^j)\Big),
\cr
d^X_\alpha \pi^i=0.
\end{cases}
\ee
The first set of equations can be viewed as evolution equations for the momentum 2-forms $\pi^i$. However, it is better to think of them as first-order evolution equations for the matrix $M^{ij}$. The matrix $M^{ij}$ then determines $X^{ij}$ via (\ref{eq-M}), which in turn determines the evolution of $\alpha^i$ via (\ref{X-alpha}). The second set of equations are the constraints that the matrix $M^{ij}$ should satisfy, relating $M^{ij}$ to $X^{ij}$. One of the important features of the connection formulation is that there are only 3 constraint equations to be satisfied, to be compared with 4 diffeomorphism constraints in the metric approach. We further discuss the evolution equations (\ref{feqs}) from the Hamiltonian viewpoint in the Appendix. 

\subsection{Partial gauge-fixing}

Let us now assume $A^i$ to be an asymptotically hyperbolic connection. As we analyse in more details in the Appendix, there is a gauge freedom of ${\rm SO}(3)$ rotations, together with diffeomorphisms. This can be used to fix a convenient gauge. For now, we just fix a part of the gauge freedom, by setting $\beta^i=0$. This can always be done by e.g. an ${\rm SO}(3)$ gauge transformation. (Geometrically, one uses parallel transport in the radial directions to identify the restrictions of $P$ to the hypersurfaces $\rho = \text{const}$ and in this gauge $\beta^i=0$; this is sometimes called ``temporal gauge'' when treating the equations as an initial value problem.)  We then write the $\alpha^i$ part of the connection in the form 
$$\alpha^i=\frac{1}{\rho}e^i+\omega^i_{(0)},$$
with
$$ e^i = e^i_{(0)} + \rho \, e^i_{(1)} + \ldots,$$
and $\omega_{(0)}$ being the Levi-Civita connection of the frame $e_{(0)}$. Note that the gauge-fixing condition $\beta^i=0$ is not compatible with the gauge used in the metric setting, see (\ref{FG}). Indeed, as we have previously computed, the connection corresponding to metric (\ref{FG}) is of the form (\ref{A-metric}), and in particular has a $d\rho$ term. The gauge $\beta^i=0$, quite convenient in calculations, will thus fail to produce a metric in the gauge (\ref{FG}). This just means that the most natural coordinate systems used in the metric and connection settings are different.

\subsection{The matrix $X$}

The derivative of the $\alpha^i$ with respect to the radial coordinate, denoted by the dot, is 
\be
\dot{\alpha}^i = -\frac{1}{\rho^2} e^i + \frac{1}{\rho} \dot{e}^i.
\ee
The spatial curvature of $\alpha$ is given by
\be
f_\alpha^i=\frac{1}{\rho^2}\frac{1}{2}\epsilon^{ijk}e^j\wedge e^k+\frac{1}{\rho}D e^i+f^i_{(0)},
\ee
where $D\equiv d^X_{\omega_{(0)}}$ is the covariant derivative along the boundary $X$ with respect to $\omega_{(0)}$ and $f^i_{(0)}: = d\omega_{(0)}^i + (1/2)\epsilon^{ijk} \omega_{(0)}^j\omega_{(0)}^k$ is the curvature of $\omega_{(0)}$. 

We now define the matrix $X^{ij}$ via
\be\label{V0}
f_\alpha^{(i}\wedge \dot{\alpha}^{j)} = X^{ij}\, {\mathcal V}_0, \qquad {\mathcal V}_0 = \frac{1}{\rho^4} d\rho\wedge \frac{1}{6}\epsilon^{ijk} e^i_{(0)}\wedge e^j_{(0)}\wedge e^k_{(0)}.
\ee
The reason for choosing this volume form to define $X^{ij}$ is that its expansion in powers of $\rho$ starts with the identity matrix
\be\label{X-exp}
X^{ij} = \delta^{ij} + \rho X^{ij}_{(1)} +\ldots.
\ee
We will compute terms in this expansion below. 

The fact that the expansion for $X^{ij}$ starts with the identity implies that also the matrix $M^{ij}$, which is algebraically expressed in terms of $X^{ij}$ via (\ref{eq-M}), to leading order is a multiple of the identity matrix. For example, for the Einstein case we have
\be
\frac{\Lambda}{3} M = \sum_n (-1)^n \left( \frac{3\Psi}{\Lambda}\right)^n,
\ee
where $\Psi$ is a trace free matrix whose expansion starts with order $\rho$ (or higher). So, we write
\be\label{m-exp}
M \sim m^{ij} = \delta^{ij} + \rho m^{ij}_{(1)} + \ldots,
\ee
 for an appropriate multiple of $M$. This expansion can then be used in field equations (\ref{feqs}). 

\subsection{The evolution equation for $m^{ij}$}

We can now write down the first of the equations in (\ref{feqs}) as an equation on the matrix $m\sim M$, where for a general theory an appropriate multiple of $M$ is taken so that the expansion for $m$ starts with the identity matrix. After some cancelations and rewriting, the equation reads
\be\label{m-eqn}
\dot{m}^{ij} \left( \frac{1}{\rho^2}\frac{1}{2}\epsilon^{jkl}e^k\wedge e^l+\frac{1}{\rho}D e^j+f^j_{(0)}\right) = 
(Dm^{ij})\wedge \left( -\frac{1}{\rho^2} e^i + \frac{1}{\rho} \dot{e}^i\right) \\ \nonumber
+ \frac{1}{\rho} (\epsilon^{ikj} m^{jl} - m^{ij} \epsilon^{jkl}) e^k  \wedge \left( -\frac{1}{\rho^2} e^l + \frac{1}{\rho} \dot{e}^l\right).
\ee
The left-had-side starts with terms of the order $1/\rho^2$, while there is a term of order $1/\rho^3$ on the right. This term cancels only when the expansion of $m$ starts with a multiple of the identity matrix. So, this equation once again tells us (\ref{m-exp}) should hold. 

\subsection{Solving for $m_{(1)}$}

We get an equation for $m_{(1)}$ by keeping terms of order $1/\rho^2$ on both sides. We get
\be
\dot{m}^{ij} \frac{1}{2}\epsilon^{jkl}e^k_{(0)}\wedge e^l_{(0)} + (\epsilon^{ikj} m^{jl}_{(1)} - m^{ij}_{(1)} \epsilon^{jkl}) e^k_{(0)}  \wedge e^l_{(0)} =0.
\ee
We now use
\be
e^i\wedge e^j = \epsilon^{ijk} \frac{1}{2} \epsilon^{klm} e^l\wedge e^m
\ee
to get
\be
0=m^{ij}_{(1)} + (\epsilon^{ikn} m^{nl}_{(1)} - m^{in}_{(1)} \epsilon^{nkl})\epsilon^{klj}= - 2m^{ij}_{(1)} + \delta^{ij} {\rm Tr}(m_{(1)}),
\ee
which implies that
\be
m_{(1)}=0.
\ee

\subsection{Solving for $m_{(2)}, m_{(3)}$}

We get $m_{(2)}$ by exactly the same logic, keeping terms of order $1/\rho$ on both sides. We get
\be
0=2m^{ij}_{(2)} + (\epsilon^{ikn} m^{nl}_{(2)} - m^{in}_{(2)} \epsilon^{nkl})\epsilon^{klj}= - m^{ij}_{(2)} + \delta^{ij} {\rm Tr}(m_{(2)}),
\ee
which implies that
\be
m_{(2)}=0.
\ee

The new phenomenon appears at the next order, where we keep $O(1)$ terms. We now get
\be
0=3m^{ij}_{(3)} + (\epsilon^{ikn} m^{nl}_{(3)} - m^{in}_{(3)} \epsilon^{nkl})\epsilon^{klj}=  \delta^{ij} {\rm Tr}(m_{(3)}).
\ee
So, in this case the evolution equation only constraints the trace part of $m_{(3)}$ to vanish. Its trace free part remains arbitrary. 

\subsection{The constraint}

The second constraint equation in (\ref{feqs}) takes the form
\be\label{constraint}
D\left( m^{ij} \left( \frac{1}{\rho^2}\frac{1}{2}\epsilon^{jkl}e^k\wedge e^l+\frac{1}{\rho}D e^j+f^j_{(0)} \right) \right)
+ \frac{1}{\rho} \epsilon^{ijk} e^j \wedge m^{kn} \left( \frac{1}{\rho}D e^n+f^n_{(0)}\right) =0.
\ee
We dropped one of the terms inside the second pair of brackets here, because it is easy to see that it is proportional to $\epsilon^{ijk} m^{jk}$, which is zero in view of the symmetry of $m$. As we already know from the considerations above, the expansion for $m$, after the leading identity matrix term, continues with a term of order $\rho^3$, which is trace free. Then the leading order term in (\ref{constraint}) is
\be
(Dm^{ij}_{(3)})\wedge \epsilon^{jkl} e^k_{(0)}\wedge e^l_{(0)}=0,
\ee
which gives a further restriction on the matrix $m$.

\subsection{Solving for the connection}

From the fact that the matrix $m\sim M$ has an expansion of the form $m = \id + \rho^3 m_{(3)}+\ldots$, together with the fact that $X$ is algebraically determined by $M$ up to scale, see (\ref{eq-M}), we can immediately deduce that in (\ref{X-exp})
\be\label{X-12}
X_{(1)}^{ij}\Big|_{tf}= X_{(2)}^{ij}\Big|_{tf}=0.
\ee
We can only conclude that the trace free parts vanish, because the trace part of $X$ is not constrained by the equation (\ref{eq-M}) (indeed $X$ only makes invariant sense up to overall scale). 

Let us now compute $X_{(1)}^{ij}$ in terms of the connection. To this end, it is convenient to write
\be
e^i = \gamma^{ij} e^j_{(0)}.
\ee
The matrix $X$ is given by
\be
X^{ij} \frac{1}{6}\epsilon^{ijk} e^i_{(0)}\wedge e^j_{(0)}\wedge e^k_{(0)} = \left( \frac{1}{2} \epsilon^{(i|kl} e^k\wedge e^l + \rho D e^{(i} + \rho^2 f^{(i}_{(0)}\right) \wedge \left( e^{j)} - \rho \dot{e}^{j)}\right).
\ee
A simple computation gives
\be\label{eq-gamma-1}
X^{ij}_{(1)} = \epsilon^{(i|kl} \gamma^{kn}_{(1)} \epsilon^{nl|j)} = \delta^{ij} {\rm Tr}(\gamma_{(1)}) - \gamma_{(1)}^{(ij)}.
\ee
It is clear that we can only conclude from (\ref{X-12}) that the trace free part of $\gamma_{(1)}$ vanishes. To determine all of it we need to do more gauge-fixing. 

\subsection{Further gauge-fixing}

We fix the gauge completely. Thus, we require that $\gamma$ be symmetric and, also, that all terms except $\gamma_{(0)}=\id$ be trace free
\be\label{gauge}
\gamma^{ij}=\gamma^{(ij)}, \qquad {\rm Tr}(\gamma_{(n)})=0, \quad n>1.
\ee
The fact that this gauge condition, together with the imposed earlier $\beta^i=0$ is an admissible gauge, is demonstrated in the Appendix. 

\subsection{Connection determined}

Having fixed the gauge, the equation (\ref{X-12}) with (\ref{eq-gamma-1}) fix 
\be\label{gamma-1}
\gamma_{(1)}^{ij}=0.
\ee

Taking (\ref{gamma-1}) into account, it is easy to compute $X_{(2)}$. Taking into account the gauge-fixing conditions (\ref{gauge}) we get
\be
X_{(2)}^{ij} = - 2\gamma_{(2)}^{ij} + f^{ij}_{(0)},
\ee
where we introduced the components of the curvature
\be\label{f-ij}
f^i_{(0)} = \frac{1}{2} f^{ij}_{(0)} \epsilon^{jkl} e^k_{(0)}\wedge e^l_{(0)}.
\ee
The matrix $f^{ij}_{(0)}$ is related to the Ricci curvature, see below, and is hence symmetric. From (\ref{X-12}) we get
\be
\gamma_{(2)}^{ij} = \frac{1}{2} \left( f^{ij}_{(0)} - \frac{1}{3} {\rm Tr}(f_{(0)}) \delta^{ij}\right).
\ee

Now, from (\ref{f-ij}) the Riemann curvature of the zeroth order metric is given by
\be
R^{ij \, mn}_{(0)} = \epsilon^{ikj}  f^{kl}_{(0)} \epsilon^{lmn}.
\ee
Hence the Ricci tensor is 
\be
R_{(0)\,ij} = R_{(0)\,imj}{}^m = f_{(0)\,ij} - \delta_{ij} {\rm Tr}(f_{(0)}).
\ee
or equivalently
\be
f^{ij}_{(0)} = R^{ij}_{(0)} - \frac{1}{2} \delta^{ij} R_{(0)}.
\ee
Hence, we can equivalently write $\gamma_{(2)}$ as
\be
\gamma_{(2)}^{ij} =\frac{1}{2}\left( R^{ij}_{(0)} - \frac{1}{3} R_{(0)} \delta^{ij}\right).
\ee

\subsection{The resulting connection}

Thus, the asymptotically hyperbolic connections satisfying field equations take the form
\be\label{A-final}
A^i=\frac{1}{\rho}\Big(e_{(0)}^i+\rho \omega_{(0)}^i+\frac{\rho^2}{2} \left( R^{ij}_{(0)} - \frac{1}{3} R_{(0)} \delta^{ij}\right) e_{(0)}^j+\rho^3 e_{(3)}^i+\cdots\Big).
\ee
This differs from the connection (\ref{A-metric}) computed from an asymptotically hyperbolic metric with the absence of a $d\rho$ term. The two connections are nonetheless gauge equivalent. We emphasise that the connection has the expansion as above for any member of our class of gravity theories, including GR. 

This expansion is the analog of the metric expansion reviewed in the Introduction. The freedom that arises in the expansion of $m^{ij}$ at order $m_{(3)}^{ij}$ translates into the freedom in $X_{(3)}^{ij}$, which in turn translates into $e_{(3)}^i$ being not completely determined. After the free part of $e_{(3)}^i$ is specified, the connection equations can be solved order by order in the expansion parameter, with no further ambiguity. 

\section{Renormalised volume and Chern-Simons functional}
\label{sec:renorm}

\subsection{The matrix of wedge products}

It is a useful exercise to compute the matrix $X^{ij}$ of curvature wedge products directly from (\ref{A-final}). We have
\begin{multline}\label{X-soln}
F^i\wedge F^j = - 2\frac{d\rho}{\rho^2} \wedge e_{(0)}^{(i} \wedge \frac{1}{2\rho^2} \epsilon^{j)kl} e_{(0)}^k \wedge e_{(0)}^l \\ 
-2\frac{d\rho}{\rho^2} \wedge e_{(0)}^{(i} \wedge ( f^{j)}_{(0)} + \epsilon^{j)kl} e_{(0)}^k \wedge(Re_{(0)})^l )\\ + 2d\rho \wedge(R e_{(0)})^{(i} \wedge \frac{1}{2\rho^2} \epsilon^{j)kl} e_{(0)}^k \wedge e_{(0)}^l + \ldots,
\end{multline}
where we denoted 
\be
(Re_{(0)})^i:= \left( R^{ij} - \frac{1}{3} R \delta^{ij}\right) e_{(0)}^j,
\ee
and dots stand for terms of order $d\rho/\rho$ and higher. These terms vanish when the trace is taken, and so will not contribute into our computation of the divergent terms below.

\subsection{Divergences}

We now compute the divergences arising in evaluating the action (\ref{action}). As in the metric context, we compute the bulk integral up to surface $\rho=\epsilon$. To perform the computation, we need to know not only the matrix of wedge products (\ref{X-soln}), but also the matrix $M$. As we have previously established, this matrix has the following expansion (in the case of GR):
\be\label{M-soln}
M = \frac{3}{\Lambda} \left( \id + O(\rho^3) \right),
\ee
where moreover the order $\rho^3$ term is tracefree. 

We now combine (\ref{X-soln}) and (\ref{M-soln}) to compute the divergent contributions. The leading term in (\ref{X-soln}) is a multiple of the identity matrix. Hence, because of the trace free character of the order $\rho^3$ term in (\ref{M-soln}), there is no contribution from that term and this is why it was not necessary to compute it. Thus, all the divergent terms in (\ref{action}) are the same as those in the functional
\be\label{bulk-bound}
\frac{3}{\Lambda}\int_\epsilon {\rm Tr}(F\wedge F) = - \frac{3}{\Lambda} S^{CS}_{X_\epsilon}[A] ,
\ee
where $S^{CS}$ is the Chern-Simons functional (\ref{CS}). The minus sign in (\ref{bulk-bound}) is due to our choice of the orientation: we have placed the asymptotic boundary in the bulk integral on the lower limit of integration.  We thus see that the boundary term necessary to remove the divergences arising in the bulk integration is just a multiple of the boundary Chern-Simons action. This gives another reason for selecting the Chern-Simons functional of the boundary connection as the boundary term in our action principle: it serves both the purpose of making the variational principle well-defined as well as renormalising the action by subtracting the divergent contributions. 

It is interesting to compute the divergences directly. Taking the trace of (\ref{X-soln}), and using the fact that the matrix in $(Re_{(0)})$ is trace free, we get
\be
\int_\epsilon {\rm Tr}(F\wedge F) = -2 \int_{\epsilon} \int_{X_\epsilon} (3 + \rho^2 {\rm Tr}(f)) + \ldots=
-\frac{2}{\epsilon^3} \int_{X_\epsilon} dv_0 - \frac{2}{\epsilon} \int_{X_\epsilon} dv_0 \,{\rm Tr}(f)+ \ldots,
\ee
where dots stand for finite parts, and $dv_0$ is the volume form for the metric with triad $e_{(0)}$. 

\subsection{Renormalised volume}

Finally, we can define the notion of renormalised volume in the connection setup. We define it as
\be\label{RV-CS}
-2 \Lambda \, RV= \lim_{\epsilon\to 0} \left( \int_\epsilon {\rm Tr}(M F\wedge F) + \frac{3}{\Lambda} S^{CS}_{X_\epsilon} \right),
\ee
where we took (\ref{volume}) into account. 

We note that the coefficient in front of $1/\epsilon$ term in our connection setup is different from that in the metric setup, see (\ref{RV}). This is because we are working in a different gauge and the metric as computed from the connection would have had the mixed $d\rho dx^i$ terms. These would change the computation of the renormalised volume above. There would be extra sub-leading terms present that would change the coefficient in front of the $1/\epsilon$ term. 

\subsection{Large gauge transformations}

We have defined renormalised volume using a multiple of the Chern--Simons functional for the renormalisation. However, as is well-known, Chern--Simons functional is not gauge invariant. It remains invariant under small gauge transformations (homotopic to the identity). Under the so-called large gauge transformations, Chern--Simons functional changes (in our normalisation) by an integral multiple of $8\pi^2$, so that $S^{CS}/4\pi$ changes by a multiple of $2\pi$. More precisely, there is a natural isomorphism $H_3(\rm{SU}(2)) \cong \mathbb{Z}$; now given a gauge transformation $g \colon X \to \rm{SU}(2)$ of the trivial bundle, write $n$ for the integer $g_*[X] \in H_3(\rm{SU}(2)) \cong \Z$, then $S^{CS}(g^*(A)) = 8\pi^2 n + S^{CS}(A)$. Thus, our renormalised volume is defined modulo addition of a constant:
\be
RV\in \mathbb{R}\quad {\rm mod} \quad \frac{3}{2\Lambda^2} 8\pi^2. 
\ee
This has interesting consequences. The Riemannian signature Einstein--Hilbert action
\be
S_{\rm EH} = -\frac{1}{16\pi G} \int \sqrt{g} (R-2\Lambda)
\ee
becomes, on Einstein metrics $R_{\mu\nu}=\Lambda g_{\mu\nu}$, a multiple of the total volume of the space
\be\label{S-EH-on-shell}
S = -\frac{\Lambda}{8\pi G} V.
\ee

In the connection setup, we want to do a path integral over connections 
\be
\int {\cal D}A \, \exp{\frac{i}{\hbar} S[A]}.
\ee
As we have described above, our action in the connection setting is just a multiple of the total volume of the space, and we should clearly use the same multiple as in (\ref{S-EH-on-shell}). Now, as we discussed, in the asymptotically hyperbolic setting we need to add a multiple of the Chern--Simons functional of the boundary connection to our bulk action, to regularise it and to make the variational principle well-defined. This means that in the asymptotically hyperbolic setting, our connection action is again only defined modulo a constant:
\be
S  \in \mathbb{R} \quad {\rm (mod)} \quad \frac{3}{8 G \Lambda} 2\pi.
\ee
If we want the path integral to be insensitive to this ambiguity we should require that the phase in $\exp{i S/\hbar}$ is an integer multiple of $2\pi$. This gives the following quantisation condition
\be
\frac{1}{8 G \hbar} = \frac{N\Lambda}{3}, \qquad N\in {\mathbb N}.
\ee
To put it differently, introducing the Planck length $l_P^2 = G\hbar$ and the Hubble length $\Lambda/3=1/L^2$ we get
\be\label{quant}
L^2 = 8 N l_P^2,
\ee
the Hubble length is quantised in units of the Planck length. We find it intriguing that our connection approach, via the setting of asymptotically hyperbolic connections, requires the cosmological constant to be quantised. 

Of course, we have derived the above quantisation condition working with asymptotically hyperbolic connections corresponding to negative scalar curvature. Observations indicate that our Universe is filled with dark energy whose equation of state is consistent with having a {\it positive} cosmological constant. If we extrapolate (\ref{quant}) to the setting of positive $\Lambda$, then, for all quantities as observed $N$ is very large $8N = L^2/l_P^2 \sim 10^{120}$, and is the number embodying the famous cosmological constant problem. 

\section{Discussion}

Four-dimensional Riemannian geometry admits an equivalent description in terms of ${\rm SO}(3)\sim{\rm SU}(2)$-connections rather than metrics. In this paper we have developed an analog of the Fefferman--Graham asymptotic expansion in the connection setting. The resulting expansion appears, at least to us, to be simpler than in the metric language, but it is possible that this is a missconception that results from living in the connection land for too long. 

Some of the new features arising in the connection setting are as follows. First, instead of solving second order ODE's for the connection directly, a much more effective procedure, as our analysis above shows, is to solve first order ODE's for the auxiliary matrix $M$ first. We have seen that to the first few orders in the expansion in powers of the radial coordinate the solution is very easy, in the sense that the leading order is the identity matrix, and then the sub- and next to sub-leading orders are identically zero. The evolution equation then only equates the trace of the order $\rho^3$ term in the expansion to zero, leaving the trace free part completely arbitrary. There is another condition, which can be viewed as an analog of the transversality, on the trace free part at order $\rho^3$. This comes from the 3 constraint equations that the ``initial data'' must satisfy. Having solved for the auxiliary matrix, one immediately obtains an analogous expansion for the matrix of curvature wedge products $X^{ij}$. From this it is very easy to solve for the expansion of the connection components. 

One of the most interesting features of the connection setting is that the two different counter terms that are required in the metric setup in the process of evaluation of the renormalised volume combine into a single Chern--Simons term. This can be seen with almost no computation, and results from the fact that the auxiliary matrix $M$ that appears in the Lagrangian has the expansion that starts with sufficiently high order so that the divergences of the volume are exactly those of the integral $\int {\rm Tr}(F\wedge F)$ of the Pontryagin density. The latter is a boundary term given by the Chern--Simons functional of the connection. An example of the computation of renormalised volume in the language of connections, together with analysis of the behaviour under the deformation of the theory, will appear as a separate publication. 

As we have described, the Chern--Simons functional serves two different purposes: it renormalises the bulk action as well as makes the variational principle well-defined. We have seen that there is a price to be paid for using $S^{CS}$ as the boundary term, as it is only invariant under small gauge transformations, while changing by multiples of $8\pi^2$ under large gauge transformations. However, we have seen that this can be used to an advantage in the sense that this ambiguity gives rise to the quantisation condition (\ref{quant}), which is not incompatible with the observed values of $\Lambda$ and $G$. 

It should also be said that the quantisation condition (\ref{quant}) is reminiscent of what happens in the setting of Riemannian signature gravity in three dimensions (with positive $\Lambda$), where the full action, not just the boundary term as here, is given by the Chern-Simons functional of a certain ${\rm SU}(2)$-connection. Similar large gauge transformation considerations show that the cosmological constant must be quantised in this setting, see e.g., \cite{Freidel:1998ua} where this fact is reviewed. 

We also note that the fact that the Chern--Simons functional renormalises the volume is analogous to the statement in \cite{Anderson} that the boundary term at infinity in the Chern--Gauss--Bonnet formula renormalises the volume. The statement in this paper is similar, except that we use the integral of the first Pontryagin density of the bundle, not the Euler density, for the renormalisation. 

\section*{Acknowledgments}

JF was supported by ERC Consolidator Grant 646649-SymplecticEinstein and FNRS Grant MIS F-4522.15. YH, KK and CS were supported by ERC Starting Grant 277570-DIGT\@. 

\section*{Appendix}
\appendix
\section{Partial gauge fixing of asymptotically hyperbolic connections}

In this Appendix we show how the gauge freedom of diffeomorphisms plus gauge transformations can be used to simplify the connection asymptotic expansion problem. 

Under an infinitesimal diffeomorphism generated by $\xi+\rho\lambda\partial_\rho\in\Gamma(T\overline{M})$ and infinitesimal gauge transformation generated by $\varphi^i\in C^\infty(\overline{M},\su(2))$ the connection (\ref{as-hyperb-conn}) transformations as
\begin{multline*}
\delta_{\xi+\rho\lambda\partial_\rho+\varphi}A^i
=\frac{1}{\rho}(\cL^X_\xi e^i_{(0)}+\epsilon^{ijk}e^j_{(0)}\varphi^k-\lambda e^i_{(0)})
\\
+\cL^X_\xi \omega^i_{(0)}+d^X\varphi^i+\epsilon^{ijk}\omega_{(0)}^j\varphi^k+\cL^X_\xi a^i+\epsilon^{ijk}a^j\varphi^k+\rho(\lambda \dot a^i+b^id^X\lambda)
\\
+\Big(\frac{1}{\rho}\iota_{\dot \xi}e^i_{(0)}+\iota_{\dot \xi}\omega^i_{(0)}+\dot \varphi^i+\iota_{\dot \xi}a^i+\iota_\xi d^X b^i+\epsilon^{ijk} b^j\varphi^k+\lambda b^i+\rho(\lambda\dot b^i+\dot \lambda b^i)\Big)d\rho
\end{multline*}
The idea now is to set $b^i$ and the anti-symmetric and trace parts of (the subleading orders of) $\langle a^i,e^j_{(0)}\rangle$ to zero by (infinitesimal) diffeomorphisms and gauge transformations. Indeed, we may impose
$$b^i+\frac{1}{\rho}\iota_{\dot \xi}e^i_{(0)}+\iota_{\dot \xi}\omega^i_{(0)}+\dot \varphi^i+\iota_{\dot \xi}a^i+\iota_\xi d^X b^i+\epsilon^{ijk} b^j\varphi^k+\lambda b^i+\rho(\lambda\dot b^i+\dot \lambda b^i)=0,$$
\begin{multline*}
\epsilon^{ijk}\langle a^j,e^k_{(0)}\rangle+\frac{2}{\rho}\varphi^i+\epsilon^{ijk}\langle d^X\varphi^j,e^k_{(0)}\rangle
\\
+\langle \omega^k_{(0)},e^k_{(0)}\rangle \varphi^i-\langle \omega^i_{(0)},e^k_{(0)}\rangle \varphi^k+\langle a^k,e^k_{(0)}\rangle\varphi^i-\langle a^i,e^k_{(0)}\rangle\varphi^k
\\
+\epsilon^{ijk}\langle \cL_\xi^Xa^j,e^k_{(0)}\rangle+\frac{1}{\rho}\epsilon^{ijk}\langle \cL^X_{\xi}e^j_{(0)},e^k_{(0)}\rangle
+\epsilon^{ijk}\langle \cL^X_{\xi}\omega_{(0)}^j,e^k_{(0)}\rangle
\\
+\rho(\lambda\epsilon^{ijk}\langle \dot a^j,e^k_{(0)}\rangle+\epsilon^{ijk}b^j\langle d^X\lambda,e^k_{(0)}\rangle)=0,
\end{multline*}
\begin{multline*}
\langle a^k,e^k_{(0)}\rangle-\frac{3}{\rho}\lambda+\rho(\lambda\langle \dot a^k,e^k_{(0)}\rangle+b^k\langle d^X\lambda,e^k_{(0)}\rangle)
\\
+\langle\cL_\xi^Xa^k,e^k_{(0)}\rangle+\frac{1}{\rho}\langle \cL^X_{\xi}e^k_{(0)},e^k_{(0)}\rangle+\langle \cL^X_{\xi}\omega_{(0)}^k,e^k_{(0)}\rangle
\\
+\langle d^X\varphi^k,e^k_{(0)}\rangle-\epsilon^{klm}\langle \omega^k_{(0)},e^l_{(0)}\rangle\varphi^m-\epsilon^{klm}\langle a^k,e^l_{(0)}\rangle\varphi^m=0,
\end{multline*}
which can be solved recursively starting with, say, $\xi_{(0)}=0$, $\varphi_{(0)}^i=0$ and $\lambda_{(0)}=0$.

We can thus partially gauge fix the diffeomorphism and gauge freedom restricting our considerations to asymptotically hyperbolic connections of the form
$$A^i=\frac{1}{\rho}e^i+\omega^i_{(0)},$$
with $e^i$ symmetric and trace 3 with respect to the frame $e^i_{(0)}$.

There still remains some diffeomorphism and gauge freedom: infinitesimal diffeomorphisms and gauge transformations generated by $\xi+\rho\lambda\partial_\rho\in\Gamma(T\overline{M})$ and $\varphi^i\in C^\infty(\overline{M},\su(2))$ satisfying
$$\frac{1}{\rho}\dot \xi^i+\dot \xi^k\langle\omega^i_{(0)},e^k_{(0)}\rangle+\dot\varphi^i+\dot \xi^k\langle a^i,e^k_{(0)}\rangle=0,$$
\begin{multline*}
\frac{2}{\rho}(\varphi^i-\varphi_{(0)}^i)+\epsilon^{ijk}\langle d^X(\varphi^j-\varphi_{(0)}^j),e^k_{(0)}\rangle+\langle \omega^k_{(0)},e^k_{(0)}\rangle(\varphi^i-\varphi_{(0)}^i)-\langle \omega^i_{(0)},e^k_{(0)}\rangle(\varphi^k-\varphi_{(0)}^k)
\\
+\langle a^k,e^i_{(0)}\rangle\varphi^k_{(0)}-\langle a^i,e^k_{(0)}\rangle\varphi^k+2\langle d^X\lambda_{(0)},e^i_{(0)}\rangle
\\
+\epsilon^{ijk}\langle \cL_\xi^Xa^j,e^k_{(0)}\rangle+\frac{1}{\rho}\epsilon^{ijk}\langle \cL^X_{\xi-\xi_{(0)}}e^j_{(0)},e^k_{(0)}\rangle
+\epsilon^{ijk}\langle \cL^X_{\xi-\xi_{(0)}}\omega_{(0)}^j,e^k_{(0)}\rangle
-\epsilon^{ijk}\langle a^j,e^l_{(0)}\rangle\langle \cL^X_{\xi_{(0)}} e^l_{(0)},e^k_{(0)}\rangle=0,
\end{multline*}
\begin{multline*}
-\frac{3}{\rho}(\lambda-\lambda_{(0)})
+\langle\cL_\xi^Xa^k,e^k_{(0)}\rangle+\frac{1}{\rho}\langle \cL^X_{\xi-\xi_{(0)}}e^k_{(0)},e^k_{(0)}\rangle+\langle \cL^X_{\xi-\xi_{(0)}}\omega_{(0)}^k,e^k_{(0)}\rangle
-\langle a^k,e^l_{(0)}\rangle\langle \cL^X_{\xi_{(0)}} e^l_{(0)},e^k_{(0)}\rangle 
\\
+\langle d^X(\varphi^k-\varphi_{(0)}^k),e^k_{(0)}\rangle-\epsilon^{klm}\langle \omega^k_{(0)},e^l_{(0)}\rangle(\varphi^m-\varphi_{(0)}^m)=0,
\end{multline*}
for arbitrary $\xi_{(0)}\in\Gamma(TX)$,$\lambda_{(0)}\in C^\infty(X)$ and $\varphi^i_{(0)}\in C^\infty(X,\su(2))$, do not alter the form of the connection. These equations can again be solved recursively leading to a residual boundary diffeomorphism, rescaling and gauge freedom.

\section{Hamiltonian formulation of the connection field equations}

We here show how to derive the 3+1 decomposition of the field equations described in \ref{subsec: Init val form} from a Hamiltonian formulation. As before, we decompose the connection $A^i$ as  
$$A^i=\alpha^i+\beta^id\rho,$$
and its curvature as
$$F_A^i=f_\alpha^i-(\dot\alpha^i-d^X_\alpha\beta^i)\wedge d\rho,$$
where $d^X_\alpha$ is the covariant exterior derivative with respect to connection $\alpha^i$ along the hypersurface $\rho=\text{const}$, and $f^i_\alpha$ is the curvature of $\alpha^i$. The matrix $X^{ij}$, defined up to scale, now reads
\[
X^{ij} {\mathcal V}= F^i \W F^j =  f_\alpha^{(i}\wedge (\dot\alpha^{j)}-d^X_\alpha\beta^{j)})\wedge d\rho.
 \]
The Lagrangian \eqref{action} thus take the following 3+1 form
\be
S[A,M,\mu] = \int_M \left( M^{ij} f_\alpha^{i} \wedge \left(  \dot\alpha^{j}-d^X_\alpha\beta^{j}\right) - \mu \, g(M) \right) \W d\rho.
\ee
where we implemented the constraint with a 3-form Lagrange multiplier $\mu \in \Lambda^3\left(X \right)$.\\
The associated momenta are:
\begin{align*}
\pi^i := \pi^i_{\alpha} = M^{ij} f^i_\alpha \;; & &
\pi^i_{\beta}=0 \;; & &
\pi^{ij}_M =0 \;; & & \pi_{\mu}=0.
\end{align*}
Note that $\dot{\alpha}$ do not appear in the right-hand-side of the first equation, thus we cannot invert the system and write $\pi_{\alpha}$ as a function of $\dot{\alpha}$. The same is true for all other momenta. This means that we have to keep track of the above equations as constraints on the phase space. The generalised Hamiltonian associated to the Lagrangian has all these constraints added with corresponding Lagrange multipliers
\begin{multline}
H[\alpha, \pi, \beta, \pi_{\beta}, M , \pi_M ,u,v,w] = \int_X -\beta^{i}\left( d^X_\alpha\pi^i \right)  + \mu \, g(M) \\ 
+ u^i \W  \left(\pi^i- M^{ij} f^j_\alpha\right) + v^i \W \pi^i_{\beta} + w^{ij} \W \pi^{ij}_M + x \pi_{\mu}
 \end{multline}
where $u$, $v$, $w$ and $x$ are Lagrange multipliers introduced to take care of the constraints on the momenta. We also used integration by parts to make explicit the role of $\beta^i$ as a Lagrange multiplier.
 We have the secondary constraints:
\begin{align*}
\dot{\pi}_{\beta}=0& \Leftrightarrow d^X_{\alpha} \pi^i =0 \\
 \dot{\pi}_{M} =0& \Leftrightarrow  \mu \frac{\pa g}{\pa M^{ij}} = u^{(i} \W f^{j)}_{\alpha} \\
 \dot{\pi}_{\mu}=0& \Leftrightarrow g\left(M \right)=0
\end{align*}
At this point, Hamilton's equations read:
\begin{align*}
&\dot{\pi}^i = -[\beta,\pi]^i + d^X_{\alpha} \left( M^{ij} u^j\right) \;;& & \dot{\alpha}^i = d^X_{\alpha} \beta^i + u^i.
\end{align*}
We can now solve for $u^i$ and obtain the first equation in (\ref{feqs}). The second secondary constraint becomes 
\begin{align}\label{apdx: sec ctr}
\mu \frac{\pa g}{\pa M^{ij}} = X^{ij} 
\end{align} 
Altogether, we have reproduced the complete equations of the theory from the Hamiltonian viewpoint. 

In principle, we should also take care of the secondary constraint $\left\{ H , \pi^i- M^{ij} f^j_\alpha \right\}=0$ together with tertiary constraints possibly arising. We do not carry out this analysis here as it is not necessary to obtain the field equations. We also note that one way to complete the Hamiltonian analysis is via an Ashtekar-like Hamiltonian formulation, which for modified theories considered here was described in \cite{Krasnov:2008zz}.

\end{document}